\documentstyle[12pt,openbib]{article}
\title{Geometrical Hyperbolic Systems for\\ General Relativity and 
Gauge Theories\cite{ded}} 
\author{Andrew Abrahams, Arlen Anderson, Yvonne Choquet-Bruhat\cite{YCBadd}\\ 
and James W. York, Jr.\\
         {\it Department of Physics and Astronomy}\\
         {\it University of North Carolina,  Chapel Hill 27599-3255 USA}}
\date{}

\def\dzero{\partial_0}
\def\dzeroh{\hat\partial_0}
\def\Boxh{\hat{\mbox{\kern-.0em\lower.3ex\hbox{$\Box$}}}}

\begin{document}
\maketitle
\vspace{-10cm}
\hfill IFP-UNC-516

\hfill TAR-UNC-053

\hfill gr-qc/9605014
\vspace{7cm}

\begin{abstract}
The evolution equations of Einstein's theory and of Maxwell's 
theory---the latter used as a simple model to illustrate the former---
are written in gauge covariant first order symmetric hyperbolic 
form with only physically natural characteristic directions and 
speeds for the dynamical variables.  Quantities representing gauge 
degrees of freedom [the spatial shift vector $\beta^{i}(t,x^{j})$ 
and the spatial scalar potential $\phi(t,x^{j})$, respectively] are 
not among the dynamical variables: the gauge and the physical
quantities in the evolution equations are effectively decoupled.  
For example, the gauge quantities could be obtained as functions 
of $(t,x^{j})$ from subsidiary equations that are not part of 
the evolution equations.  Propagation 
of certain (``radiative'') dynamical variables along the physical 
light cone is gauge invariant while the remaining dynamical 
variables are dragged along the axes orthogonal 
to the spacelike time slices by the propagating variables.  We 
obtain these results by $(1)$ taking a further time derivative 
of the equation of motion of the canonical momentum, and $(2)$ 
adding a covariant spatial derivative of the momentum constraints 
of general relativity (Lagrange multiplier $\beta^{i}$) or of
the Gauss's law constraint of electromagnetism (Lagrange 
multiplier $\phi$).  General relativity also requires a 
harmonic time slicing condition or a specific generalization 
of it that brings in the Hamiltonian constraint when we 
pass to first order symmetric form.  
The dynamically propagating gravity fields straightforwardly
determine the ``electric'' or ``tidal'' parts of the Riemann tensor.

{\bf PACS \#: } 04.20.-q, 97.60.Lf
\end{abstract}
\newpage

\section{Introduction}
\indent We examine the Cauchy problem \cite{CBY79} for general relativity 
as the time history of the two fundamental forms of the geometry
of a spacelike hypersurface, its metric ${\bf {\bar g}}$ and its 
extrinsic curvature ${\bf K}$.  By using a $3+1$ decomposition of 
the Riemann and Ricci tensors, we split the Einstein equations 
into initial data constraints---equations containing 
only ${\bf \bar g} $, ${\bf K}$, and their space derivatives---and evolution 
equations giving the time derivatives of ${\bf \bar g}$ and  ${\bf K}$ 
in terms of space derivatives of these quantities and also of 
the lapse and shift, that is, of the variables that fix the proper 
time separation between leaves of the foliation of spacetime 
and the ``time lines'' threading the foliation \cite{Lic,YFB,Yor79}.  
The constraints 
can be posed and solved as an elliptic system by known methods. 
(See, for example \cite{CBY79,Yor79,OMY}.)  However, the equations of 
evolution of ${\bf \bar g} $ 
and  ${\bf K }$ are in essence the spatially covariant
canonical Hamiltonian \cite{ADMDir} equations with first order time
derivatives and second order space derivatives. 
Casting these equations into hyperbolic form
\cite{CBR,FritReu,BMSS,Fre,FiMa} is a problem important for
theoretical analysis and for practical applications.  However,
the procedure used may entail a loss of spatial coordinate
covariance, the use of non-geometrical variables, or
the appearance of unphysical characteristics, that is,
directions which are neither along the light cone
nor orthogonal to the time slices.

All of these difficulties can be overcome, as we show, by
obtaining first a nonlinear wave equation for the evolution
of ${\bf K}$.  The evolution of the spatial metric
${\bf \bar g}$ is just its dragging by ${\bf K}$ along
the axis orthogonal to the spacelike time slices.

One method for obtaining this result requires choosing the 
time slicing by specifying the mean (extrinsic) curvature of
 the slices.  It leads to a mixed elliptic-hyperbolic system 
for which we have recently proven local in time, global in 
space, existence theorems in the cases of compact
\cite{CBY95} or 
asymptotically flat slices \cite{ChKl,CBY95,CBY96}.

In this paper, we concentrate on a second method, which relies 
on a harmonic time slicing condition or a certain class of 
generalizations of it \cite{CBY95,prl,CBY96,AY96}.  
We obtain equations of motion 
equivalent to a covariant first order symmetric hyperbolic 
system with only physical characteristics.  We construct 
this system explicitly.  The only quantities propagated along 
the light cone are {\it curvatures}.  The space 
coordinates and the shift vector are arbitrary.  In this 
sense, the system is gauge covariant.

\section{Electromagnetism}

Maxwell's theory provides a simple example for illustrating the method we
use in treating Einstein's theory.  
The idea is to work with the dynamical or 3+1 form of 
the theory, that is, essentially the canonical form.  
By using a further time and a further
space derivative in a definite way, one proceeds 
to construct a physically natural
second order wave equation for the canonical 
momentum, in this case the electric field.
The wave equation for the electric field can 
be written in a first order symmetric
hyperbolic form (``FOSH" form) that is also flux conservative.  Furthermore,
this procedure maintains gauge covariance and produces a FOSH system that has 
{\it only} the physically natural characteristic directions given by null generators
of the light cone and timelike vectors
orthogonal to the spacelike time slices.  We refer
to these properties as ``simple physical characteristics.''
Correspondingly, one obtains ``characteristic fields''
that propagate along the light cone.
By treating this simple case first, many steps
in dealing with the technically more intricate case of general relativity
can be abbreviated because they follow the same pattern.

Consider Maxwell's theory on flat spacelike slices of flat Minkowski spacetime
with 
\begin{equation}
ds^2 = -dt^2 +g_{ij} dx^i dx^j
\end{equation}
and prescribed current $j_\mu (t,x^j) = (- \rho, j_i)$ such that 
$\nabla^\mu j_\mu = \dzero \rho + \bar \nabla ^i j_i = 0$
(an overbar denotes a spatial tensor or operator).
Using $F_{\mu \nu} = 2 \nabla_{[ \mu}A_{\nu ]}$ and
and $A_\mu=(-\phi,A_i)$, we define $R_\mu \equiv \nabla^\nu F_{\mu \nu}$,
which leads to the identities
\begin{equation}
R_0 \equiv - \bar \nabla ^j E_j,   
\end{equation}
\begin{equation}
R_i \equiv -\dzero E_i + \bar\nabla^jF_{ij} .
\end{equation}
The standard form of the Maxwell equations can be 
written as
$$
R_0 + 4 \pi \rho = 0,
\eqno{(4C)}
$$
$$
R_i - 4 \pi j_i = 0 
,
\eqno{(4E)}
$$
$$
\dzero A_i = -E_i - \bar \nabla _i \phi,
\eqno{(4D)}
$$
where $\bar \nabla _i \phi = \partial_i \phi$. ( C denotes ``constraint,''
E ``evolution,'' and D ``definition.'')
To construct the new 3+1 system, form the spatial one-form
\setcounter{equation}{4}
\begin{equation}
\Omega_i \equiv \dzero R_i - \bar \nabla_i R_0
\end{equation}
and
use (4D) and the definition of $F_{ij}$ to
obtain 
\begin{equation}
\Omega_i \equiv (-\dzero \dzero + \bar \Delta) E_i \equiv \Box E_i,
\end{equation}
where $\bar \Delta \equiv \bar \nabla^k\bar\nabla_k$. While (6)
is an identity, the Maxwell equations give us a new statement of
the evolution equation,
\begin{equation}
\Omega_i = 4 \pi (\dzero j_i + \bar\nabla_i\rho) .
\end{equation}
The new form of the Maxwell system is therefore
$$
\bar \nabla^i E_i = 4 \pi \rho,
\eqno{(8C)}
$$
$$
\Box E_i = 4 \pi (\dzero j_i + \bar\nabla_i \rho),
\eqno{(8W)}
$$
$$
\dzero A_i = -E_i - \bar\nabla_i \phi
,
\eqno{(4D)}
$$
where W denotes ``wave equation,'' and the scalar potential 
$\phi = \phi(t,x^j)$ is an arbitrary gauge field that can be
regarded for the present as prescribed.

The initial data (Cauchy data) for the new system on a spacelike
slice $\Sigma_0\ (t=0)$ are (a) $A_i$ given freely, (b) $E_i$ such that
(8C) holds for given $\rho$, and (c) $\dzero E_i$ such that (4E) holds
($R_i-4\pi j_i = 0$) for given $j_i$.  In other terms, the extra time
derivative used in constructing $\Omega_i$ requires that we give as
further Cauchy data $\dzero E_i$ satisfying on $\Sigma_0$ the usual
Maxwell equation of motion, which is (3) combined with (4E).

For any given $\phi(t,x^j)$, the new system (8W), (4D)
forms a quasi-diagonal strictly hyperbolic system in the unknowns 
$E_i$ and $A_i$.
Its characteristics at a point are the light cone and the time axis $\dzero$,
which lies inside the cone.  The cone therefore determines the propagation.
The new system in its present form implies that $E_i$ propagates at 
light speed along the cone while $A_i$ is dragged
along the time axis.  

Before passing to FOSH form, let us show that the new system
is equivalent to the standard Maxwell equations: Assume (7)
has been solved.  Then
\begin{equation}
\bar\nabla^i \Omega_i = 4 \pi \Box \rho,
\end{equation}
where current conservation, $\nabla^\mu j_\mu = 0$, has been used. 
Therefore, using (6) we find
\begin{equation}
\Box ( \bar\nabla^i E_i - 4\pi \rho) = 0.
\end{equation}

But $\bar\nabla^iE_i - 4\pi \rho=0$ ($R_0+4\pi\rho=0$)
on $\Sigma_0$ by hypothesis,
and from the ``Bianchi identity'' $\nabla^\mu R_\mu=0$ (i.e. $\nabla^\mu
\nabla^\nu F_{\mu\nu}=0$)
and current conservation $\nabla^\mu j_\mu=0$, we have
\begin{equation}
\dzero(\bar\nabla^iE_i-4\pi\rho)=\bar\nabla^i(-R_i+4\pi j_i)=0
\end{equation}
on $\Sigma_0$.
In the last equality we have used the initial data condition (c) 
that $R_i-4\pi j_i=0$ on $\Sigma_0$.
Therefore, the Cauchy data pertaining to (10) are zero and
it follows that the Maxwell constraint (4C) holds on 
$\Sigma_0 \times {\rm I}$, where ${\rm I}$ is an interval of
time. (Mathematically rigorous statements of results 
in this paper can be given in terms of 
Sobolev spaces $H_s$, or weighted Sobolev spaces
$H_{s,\delta}$ for asymptotically flat time slices.)

Now (5) and (7) together 
with the result just obtained that the constraint (4C) holds on 
$\Sigma_0 \times{\rm I}$ show
that
\begin{equation}
\dzero(R_i - 4 \pi j_i) =0
\end{equation}
on $\Sigma_0 \times{\rm I}$.
But again $R_i-4\pi j_i = 0$ on $\Sigma_0$, so it must
remain zero on $\Sigma_0 \times{\rm I}$, which is what we
wanted to prove. 

In passing to the FOSH form of the Maxwell equations, we will leave the
constraint (8C) unchanged; it can be converted to an
elliptic equation that needs to be solved only on $\Sigma_0$.
Also, the other initial-value data will be given as above.

We consider here the vacuum case
$\rho = j_i = 0$.  Defining $G_{0 i}=\dzero E_i$ and
$G_{ji}=\bar\nabla_jE_i$, we write the first order
system corresponding to (4D) and (8W) as
$$
\dzero A_i = -E_i -\bar\nabla_i \phi ,
\eqno{(4D)}
$$
$$
\dzero E_i = G_{0 i},
\eqno{(13D)}
$$
$$
\dzero G_{0i}-\bar\nabla^jG_{ji}=0,
\eqno{(13W)}
$$
$$
\dzero G_{ji}-\bar\nabla_jG_{0i}=0,
\eqno{(13A)}
$$
where A=``adjoint'', an equation obtained by
using the trivial commutator $[\dzero,\bar\nabla_i]=0$.
(It is already clear that this system has
``simple physical characteristics'' as previously
defined.)

Define the transpose ${\bf u}^T$ of a column matrix ${\bf u}$
by 
\setcounter{equation}{13}
\begin{equation}
{\bf u}^T = \left( {\bf A, ~ E, ~ G_0, ~ G_1, ~ G_2, ~ G_3} \right) ,
\end{equation}
where the final index ``i'' has been omitted for each entry.
The above equations (4D)-(13A) have the form
\begin{equation}
\dzero {\bf u} + {\bf B}^j \bar\nabla_j {\bf u} +{\bf C} = 0,
\end{equation}
where ${\bf B}^j\ (j=1,2,3)$ is an $n \times n$
square matrix and $C$ is an $n\times 1$ column vector, with $n$ 
the number of unknowns.  (With the abbreviated
notation (14), $n = 6$.)  The matrices ${\bf B}^j$ and vector ${\bf C}$
depend on $(t,x^i)$ in the linear case and on $({\bf u};t,x^i)$
in the quasi-linear case.  Equations (15) are hyperbolic and
{\it symmetrizable} if there is an $n \times n$
symmetric positive definite matrix ${\bf M}^0$ such that 
${\bf M}^0 {\bf B}^j \equiv {\bf M}^j$ is symmetric. 
(It may be that these properties hold only
in some restricted region of spacetime 
and, in the quasi-linear case, only in a functional neighborhood
of a particular ${\bf u}$).  If we carry out this
operation, we obtain an explicitly FOSH form,
\begin{equation}
{\bf M}^\alpha \nabla_\alpha {\bf u}+ {\bf M}^0 {\bf C} =
{\bf M}^0 \dzero {\bf u}+{\bf M}^j \bar\nabla_j{\bf u}+{\bf M}^0{\bf C}=0.
\end{equation}
{}From the FOSH form, one can construct ``energy norms'' and ``energy
inequalities'' that lead, by the methods of functional analysis,
to precise results on the existence and uniqueness of solutions,
and global estimates of the time of existence 
\cite{FiMa}.

In the present case, the ${\bf B}^j$'s are not symmetric
but the symmetrizer is
\begin{equation}
{\bf M}^0=
\left(\matrix{{\bf 1}_{3\times3}& {\bf 0}_{3\times3} \cr 
	     {\bf 0}_{3\times3}& {\bf g}^{-1}_{3\times3}}
\right)
\end{equation}
where $(g^{-1})=g^{ij}$. The matrix ${\bf M}^0$ is positive definite
if and only if 
$g_{ij}$ is properly Riemannian.  (This is important in the case of gravity,
where $g_{ij}$ is an unknown.  There, the
symmetrizer will be essentially two copies of ${\bf M}^0$ centered on
the diagonal of a simple larger matrix.)  We verify easily that the
${\bf M}^0{\bf B}^j$'s are symmetric.

It is clear from the form of (4D)-(13A)
that $A_i$ and $E_i$ are fields
associated with evolution along the  the time axis.  There
are also fields propagating at light speed.  
These ``characteristic fields'' can be found by returning  
to the symmetrizable form (15).  Then, to exhibit one example
of such fields explicitly,  we choose a space direction
at an event corresponding, for example, to j=1. 
This means the direction $\pm {\bf n_1}$, orthogonal
in space to the $x^1={\rm constant}$
two-surface passing through the event and lying in the $t={\rm constant}$
slice containing the event.  We solve the eigenvalue problem associated with
${\bf B}^1$ and diagonalize it with an appropriate matrix
${\bf T}$ by a similarity transformation
\begin{equation}
{\bf T} {\bf B}^1{\bf T}^{-1} = {\bf D}^1 \nonumber \\ ,
\end{equation}
where ${\bf D}^1$ is diagonal.
The non-zero eigenvalues of ${\bf B}^1$ give the speeds
of the ``characteristic fields'' with respect to
observers at rest in the time slices, whose four-velocities
are parallel to the orthogonal time axis $\dzero$
(``Eulerian observers'').  By construction, for the choice of
direction $\pm {\bf n_1}$, these characterisitic fields
include {\it only} those which propagate in the local
spacetime two-plane defined by $\dzero$ and $\pm {\bf n_1}$
at the given event.  As we change the direction $\pm {\bf n_1}$,
we move around the locally isotropic null cone at the event
and capture {\it all} the characteristic fields.

The non-zero eigenvalues of ${\bf B}^1$ are $\pm \alpha^{-1}$, 
where $\alpha=(g^{11})^{-1/2}$
is the spatial ``lapse function'' corresponding to the chosen
direction.  The quantity 
$\alpha^{-1}$, of course, {\it is} the speed of light because,
in this example, the time is proper but the spatial displacements
are not.  (In local Cartesian coordinates we obtain, of course,
$\pm 1$.)

The characteristic fields are obtained from
\begin{equation}
{\bf T} {\bf u} = {\bf u}_{\rm char}.
\end{equation}
They are found to be 
\begin{equation}
{1 \over \sqrt{2}} \alpha
\left[
G_{0 i} \pm \alpha^{-1} (G_{1 i}-v^a G_{ai}) \right],
\end{equation}
where $v^a=-g^{1a}(g^{11})^{-1}$ with $a=2,3$ is the
``shift vector'' in space corresponding to the $x^{1}=$constant
two-surface whose
``space lapse function'' is $\alpha=(g^{11})^{-1/2}$.
Substituting
into (20) the definitions of $G_{0i}$ and $G_{ji}$,
we see that the characteristic fields in this example
are the {\it differential}
electric field along the advanced and retarded ``$\pm {\bf n_1}$- branches''
of the generators of the null cones.  That is, the
fields (20) are
\begin{equation}
{1 \over \sqrt{2}} \alpha
\left[
\dzero \pm \alpha^{-1} (\bar\nabla_1-v^a \bar\nabla_a)) \right] E_i.
\end{equation}
The operator in rectangular brackets (times $1/\sqrt{2}$)
gives the geometrical form of the usual advanced and
retarded time derivatives defined with respect to
the chosen local spacetime two-plane
at the given event.  Knowing the characteristic fields
facilitates imposing physically consistent boundary
conditions\cite{AY96}, in this case, boundary conditions
compatible with causality.  As examples, 
consider a ``no outgoing waves'' condition at the
future event horizon of a black hole or a  ``no incoming
waves'' condition in the vacuum far-field of 
electromagnetic source switched on a finite time in the past.

The final points we wish to emphasize concern the facts that (1) no particular
gauge condition was imposed on $A_i$ to obtain our physical FOSH
system, and (2) the scalar potential (``gauge field'') 
$A_0 = -\phi$ played an entirely passive role.  It could
either be {\it given} or {\it obtained} (say) by a {\it supplementary}
(non-Maxwell) equation, as a function of $(t, x^i)$. Any 
mathematically consistent method of obtaining $\phi (t,x^i)$ can
be used without affecting the features of 
the Maxwell equations {\it per se} in the form in which we have
given them above.  The ``gauge sector'' and the ``physical sector''
have thus been separated to this extent.  Our FOSH system and its
simple physical characteristics are maintained in the face
of complete gauge covariance.  (A similar
statement holds for $A_0$ in the Yang-Mills equations \cite{prl} 
and for the ``shift-vector'' of the Einstein equations---see below.)

\section{Gravitation}

In Einstein's theory it is convenient to write the metric as
\begin{equation}
ds^2 = -N^2 (\theta^0)^2 +g_{ij} \theta^i \theta^j,
\end{equation}
with $\theta^0 = dt$ and $\theta^i = dx^i+\beta^i dt$,
where $\beta^i$ is the shift vector.
The cobasis $\theta^\alpha$ satisfies
\begin{equation}
d\theta^\alpha = - {1 \over 2} C_{\beta \gamma}^\alpha 
\theta^\beta\wedge \theta^\gamma
\end{equation}
with $C_{0 j}^i=-C_{j 0}^i = \partial_j \beta^i$ and
all other structure coefficients zero.  The corresponding vector
basis is given by $e_0 = \partial_t - \beta^j\partial_j$
and $e_i =\partial_i$ where $\partial_t = \partial/\partial t$,
$\partial_i = \partial/\partial x^i$, and the action of
$e_0$ on space scalars is the  Pfaffian or
convective derivative, 
$e_0[f]=\dzero f = \partial_t f - \beta^j \partial_j f$.

We shall assume throughout that the lapse function $N>0$
and the space metric $\bar {\bf g}$ on $\Sigma_t$ is properly
Riemannian.   Note that $\bar g_{ij}=g_{ij}$ and $\bar g^{ij}=g^{ij}$. 

The spacetime connection one-forms are defined by \cite{CBDW,Yor79}
\begin{equation}
\omega^\alpha_{\beta \gamma}=\Gamma^\alpha_{\beta \gamma}+
g^{\alpha \delta} C_{\delta(\beta}^\epsilon g_{\gamma)\epsilon}  
-{1 \over 2} C_{\beta \gamma}^\alpha = \omega_{(\beta \gamma)}^\alpha 
+\omega_{[\beta \gamma]} ^\alpha,
\end{equation}
where $(\beta \gamma)={1\over 2}(\beta \gamma +\gamma \beta)$,
$[\beta \gamma] = {1\over 2}(\beta \gamma- \gamma \beta)$, and
$\Gamma$ denotes a Christoffel symbol.  These connection forms
are written out in \cite{CBY96}.
In particular, the extrinsic curvature (second fundamental tensor) of
the space slices is given by
\begin{equation}
K_{ij} = -N \omega^0_{ij} \equiv -{1\over 2}N^{-1} \dzeroh g_{ij} ,
\end{equation}
where we define for any $t-$dependent space tensor ${\bf T}$ another 
such tensor $\dzeroh {\bf T}$
of the same type by setting
\begin{equation}
\dzeroh = {\partial \over \partial t} - {\cal L}_\beta,
\end{equation}
where
${\cal L}_\beta$ is the Lie derivative on $\Sigma_t$ with respect
to the vector $\beta$. 
Note that $\dzeroh$ and $\partial_i$
commute: $[\dzeroh,\partial_i]{\bf T} =0$.  

We have chosen in this paper to work with $K_{ij}$ rather
than the usual canonical momentum,  $(\det \bar g)^{1/2}(H  g^{ij}-K^{ij})$,
conjugate to $ g_{ij}$, where $H = K^i\mathstrut_i$ denotes
the mean curvature. This is a matter of choice, not of
principle.  Indeed, our general procedure can be applied to other
choices of canonical or geometric variables. (We 
note that $K_{ij}$ is conjugate to $(\det \bar g)^{1/2} g^{ij}$.)

We regard the dynamics of general relativity as being given by
the time dependence of the first and second fundamental forms
of the foliation $\Sigma_t$.  It is therefore appropriate to
use $\dzeroh$ as our time derivative and $\bar \nabla_i$, the spatial 
covariant derivative, for spatial 
derivatives ($\omega^i_{jk} = \Gamma^i_{jk}=\bar \Gamma^i_{jk}$).
An advantage of using
$\dzeroh$
is that it is always timelike because 
it is orthogonal to the spacelike slices $\Sigma_t$.  We recall
that $\partial/\partial t$ is not necessarily timelike
in Einstein's theory so that using it to define hyperbolicity can be
confusing.  Unlike the flat-spacetime Maxwell theory on
flat, spacelike hypersurfaces, here we have the non-trivial
commutator
\begin{equation}
[\dzeroh, \bar\nabla_i] u_j=
Nu_k\left[2\bar\nabla_{(i}K_{j)}\mathstrut^k-\bar\nabla^kK_{ij}
+ 2 a_{(i}K_{j)}\mathstrut^k-a^k K_{ij}\right],
\end{equation}
where $u_j$ is a spatial one-form and 
$a_i=\partial_i \log N = \bar\nabla_i \log N$
is the acceleration one-form of the ``Eulerian'' observers
at rest in the time slices.  The commutator on higher-rank tensors
is similar to (27), with corresponding terms for each
further index. 

It is necessary to relate the spacetime and space
Riemann tensor components as well as the
corresponding Ricci tensor components. We use the
conventions in \cite{MTW} and the $\theta^\alpha$ cobasis.
The Riemann tensor components are given by
\begin{eqnarray}
R_{ijkl}&=&\bar R_{ijkl}+2K_{i[k}K_{l]j},
\\
R_{0ijk}&=&2N\bar\nabla_{[j}K_{k]i},
\\
R_{0i0j}&=&N(\dzeroh K_{ij} +N K_{ik}K^k\mathstrut_j+
\bar\nabla_i\bar\nabla_j N), 
\end{eqnarray}
where we note that
\begin{equation}
N \bar\nabla_i\bar\nabla_jN=N^2(\bar\nabla_ia_j+a_ia_j),
\end{equation}
and, in three dimensions,
\begin{equation}
\bar R_{ijkl}=2g_{i[k}\bar R_{l]j}+2g_{j[l}\bar R_{k]i}
+\bar R g_{i[l}g_{k]j}.
\end{equation}
The latter relation proves to be very important
in our construction of a FOSH system for the
Einstein equations in (3+1) dimensions.

The Ricci curvature $R_{\alpha \beta} = 
R^\sigma \mathstrut_{\alpha \sigma \beta}$
is given in (3+1) form by the identities
\begin{eqnarray}
R_{ij}&=&\bar R_{ij}-N^{-1} \dzeroh K_{ij}+H K_{ij}
	-2K_{ik}K^k\mathstrut_j-\bar\nabla_i a_j -a_ia_j,\\
R_{0i}&=&N\bar\nabla^j(Hg_{ij}-K_{ij}),\\
R_{00}&=&N\bar\Delta N-N^2 K_{mk}K^{mk}+N\dzero H .
\end{eqnarray}
The Einstein equations, $G_{\alpha \beta} = \kappa T_{\alpha \beta}$,
 can be written as 
$R_{\alpha \beta} = \kappa \rho_{\alpha \beta}$,
where $\kappa= 8\pi$ if $G=c=1$, and, in (3+1) dimensions,
$\rho _{\alpha \beta}= T_{\alpha \beta} - 
{1\over 2} g_{\alpha \beta}T^\sigma\mathstrut_\sigma$.
In this paper we will take $T_{\alpha \beta}=0$ but indicate
at certain stages what the matter terms are.
(The presence of matter fields indeed causes no difficulties
in a complete hyperbolic system for gravity plus matter if
these fields themselves have well-posed Cauchy problems
in flat spacetime and, as a sufficient condition, are
minimally coupled to gravity.)

We recognize (33) as the evolution equation 
E, (34) as the ``momentum constraints'' $C_i$,
and from the definition 
$G_{\alpha \beta} = R_{\alpha \beta}- {1\over 2} R g_{\alpha \beta}$
of the Einstein tensor, we see that the ``Hamiltonian constraint'' $C_0$
is defined by 
\begin{equation}
G^0\mathstrut_0={1\over2}(R^0\mathstrut_0-R^i\mathstrut_i)
={1\over 2}( K_{mk}K^{mk}-H^2-\bar R).
\end{equation}
The constraint equations $C_i$ and $C_0$ can be posed
and solved as an elliptic problem and
will not be discussed further here.  Therefore,  in analogy
to (4C), (4E), and (4D), we have
$$
R_{0i}=0, ~~~~~~~~~~~~
{1\over2}(R^0\mathstrut_0-R^i\mathstrut_i)=0,
\eqno{(37C)}
$$
$$
R_{ij}=0,
\eqno{(37E)}
$$
$$\dzeroh g_{ij}=-2N K_{ij}.  \eqno{(37D)}$$
\setcounter{equation}{37}

To construct the new (3+1) system, form the symmetric
spatial tensor
\begin{equation}
\Omega_{ij}\equiv \dzeroh R_{ij}-2\bar\nabla_{(i}R_{j)0},
\end{equation}
and work it out using the relations above and
\begin{eqnarray}
\dzeroh \bar R_{ij} &=& \bar\nabla_k (\dzeroh \bar \Gamma^k_{ij}
) -  \bar\nabla_j (\dzeroh \bar \Gamma^k_{ik}),
\\
\dzeroh \bar \Gamma^k_{ij}&=&- 2 \bar\nabla_{(i}(N K_{j)}\mathstrut^k)
		+ \bar\nabla^k(N K_{ij}),
\end{eqnarray}
to obtain, in analogy to (6), the identity
\begin{equation}
\Omega_{ij}\equiv N {\Boxh} K_{ij}+J_{ij}+S_{ij},
\end{equation}
where $\Boxh \equiv -(N^{-1}\dzeroh )^2 +\bar\Delta$.
This wave operator was denoted ``$\Box$'' in \cite{prl};
but here we reserve
that standard symbol for $g^{\alpha \beta} \nabla_\alpha \nabla_\beta$,
where $\nabla_\alpha$ is the usual spacetime covariant derivative
operator.  In \cite{CBY96} and \cite{CBY95}, slightly
different wave operators were used, resulting in slightly
different counterparts for $J_{ij}$.
These wave operators all have the same principal symbol.

With the conventions above, to which we adhere in this
paper, the explicit form of the
``source'' $J_{ij}$ is given in \cite{prl}; below,
it will be written out explicitly for our FOSH system.
For the present, we note that $J_{ij}$ contains no derivatives
of $K_{ij}$ higher than the first; and that it contains
$\bar R_{ijkl}$ and $\bar R_{ij}$ with the former
subject to elimination in favor of the latter using (32).
The contribution of matter to (41), before the
elimination of $\bar R_{ij}$ to be carried out later in
obtaining FOSH form, is found by noting that Einstein's
equations assert
\begin{equation}
\Omega_{ij} = \kappa [\dzeroh \rho_{ij} - 2 \bar\nabla_{(i}\rho_{j)0}].
\end{equation}

The ``slicing term'' $S_{ij}$ is given by
\begin{equation}
S_{ij}=-N^{-1} \bar\nabla_i \bar\nabla_j(\dzero N + N^2 H).
\end{equation}
Because of the presence of the $\bar\nabla_i \bar\nabla_j H$ term in
$S_{ij}$, the principal part of the complete second-order operator
acting on $K_{ij}$ in (41) is not a wave operator.  However, this
problem (which has no counterpart in Maxwell's theory) can be
eliminated by using either the {\it simple harmonic slicing}
condition ($g^{\alpha \beta}\omega^0_{\alpha\beta}=0$)
\begin{equation}
\dzero N + N^2 H=0,
\end{equation}
or a {\it generalized harmonic slicing} condition \cite{prl}
\begin{equation}
\dzero N + N^2 H=Nf(t,x^j),
\end{equation}
where $f(t,x^j)$ is an arbitrary given function analogous
to a ``gauge source''\cite{Fre}.  Although we shall,
for simplicity, take $f=0$ in most of the following discussion, a
choice of $f\ne 0$ causes no material changes.  In any event, we now
have, from (41) and (45),
\begin{equation}
\Omega_{ij}\equiv N {\Boxh} K_{ij}+J_{ij}-\bar\nabla_i \bar\nabla_j f
-2 a_{(i} \bar\nabla_{j)} f -f(\bar\nabla_i a_j + a_i a_j),
\end{equation}
which is a second-order nonlinear wave operator on $K_{ij}$ for all
time slicings in the class (45). (Recall $a_i=\bar\nabla_i \log N$.)

[Another way to handle $S_{ij}$ is to prescribe the mean curvature:
set $H=h(t,x)$, a known function.  This leads to a mixed elliptic-hyperbolic
representation of the Einstein evolution equations which we will not
pursue here.  See \cite{CBY96,CBY95}.]

The new Einstein system of equations of motion is given by (46), (45),
and $\dzeroh g_{ij}=-2NK_{ij}$.  The appropriate Cauchy data on an 
initial slice $\Sigma_0$ ($t=0$) are ${\bf \bar g},\ {\bf K},\ {\rm N}$,
and $\dzeroh {\bf K}$.  We require ${\bf \bar g}$ and ${\bf K}$ to
satisfy the usual momentum and Hamiltonian constraints and $\dzeroh {\bf
K}$ to be given by $R_{ij}=0$ in (33) in the vacuum case:
that is, we require that all the Einstein equations $G_{\alpha\beta}=0$
hold on $\Sigma_0$.  The lapse function must satisfy $N>0$.  Note
that the shift $\beta^i$,  which we 
may regard at present as being
known in the form $\beta^i(t,x^j)$, is playing a passive role
completely analogous to the scalar potential $\phi$ in our treatment of
the Maxwell equations.  The shift has been absorbed into
the operator $\dzeroh$.

To show that the new system is indeed solvable, before showing that it is
equivalent to the Einstein equations on $\Sigma_0 \times I$, we shall reduce
it to the form (8W) and (4D) obtained for the Maxwell equations.  To
accomplish this, we use the expression for $H$ in terms of $g_{ij}$ and
its derivatives [cf. (37D)] to write (45) in the form
\begin{equation}
\dzero \log [N(\det \bar g)^{-1/2}]=f(t,x^j),
\end{equation}
from which $N$ can be obtained in terms of the spatial metric and the known
functions $\beta^i$ and $f$, called ``algebraic gauge'' in \cite{CBY96,CBY95}. 
(Compare \cite{CBR} in the case $\beta^i=0$
and $f=0$ and \cite{CBY96,CBY95} in the case $f=0$.)  The new system now 
acquires a reduced form
\begin{eqnarray}
{\Boxh} K_{ij} &=& -N^{-1}[J_{ij}-\bar\nabla_i \bar\nabla_j f
-2 a_{(i} \bar\nabla_{j)} f -f(\bar\nabla_i a_j + a_i a_j)], \\
\dzeroh g_{ij} &=&-2N K_{ij},
\end{eqnarray}
where N is understood to be determined by (47).  These equations, 
analogous to the Maxwell equations (8W) and (4D), are a quasi-diagonal
system whose characteristics are the light cone and the $\dzero$ axis,
which lies inside the cone.  This form of the equations implies that
$K_{ij}$ propagates along the light cone at unit speed while $g_{ij}$
is dragged by $K_{ij}$ along the $\dzero$ axis.  Since
equivalence to Einstein's theory has not yet been proven and
therefore cannot be used, the Ricci curvatures occuring in $J_{ij}$ \cite{prl}
are understood by their definition in terms of the metric $g_{ij}$
and its derivatives, up to second order.
Using (49) in (48), we obtain a third order strictly hyperbolic equation
for $(\det \bar g)^{-1/2} g_{ij}$. A 
theorem of Leray\cite{leray} then implies the temporally local
existence of solutions in spatially local Sobolev spaces. In practice,
we see that we can regard (45), (46), and $\dzeroh g_{ij}=-2N K_{ij}$
as a system to be solved simultaneously.

It is next required that we show the equivalence of the above system
to the standard Einstein equations.  This proof is transparent especially
when the actions of the operators $\dzeroh$
and $\bar\nabla_i$ on spatial scalars, one-forms, 
and covariant symmetric tensors are related to those
of $\nabla_0$ and $\nabla_i$ on these same objects.  Such relations follow 
from the definitions of the operators and the values of the connection
one-forms $\omega^\alpha_{\beta\gamma}$.

Assume that the reduced system defined by (45) or (47), as well
as (48) and (49), has been solved, so that the lapse and the
metric are known on $\Sigma_0\times {\rm I}$, where ${\rm I}$ is an 
interval of time.  We must prove that given the
Cauchy data for the reduced system, i.e., that the Einstein
equations hold initially and that $N>0$ is given 
initially in accord with (47), then (38) [and (42)], supplemented by the 
twice-contracted Bianchi identities [and the covariant conservation
law $\nabla^\beta T_{\alpha \beta}=0$], implies that 
they hold on $\Sigma_0\times {\rm I}$.  It is necessary that we have already
found (i.e. proven the existence) of the lapse and the metric so
that (38) becomes a definite equation for the Ricci curvature, and
we can raise and lower indices as necessary.  In vacuum,
\begin{equation}
\Omega_{ij} \equiv \dzeroh R_{ij}-2\bar\nabla_{(i}R_{j)0}
\equiv \nabla_0R_{ij}-2\nabla_{(i}G_{j)0} +({\rm l.h. E.})_0=0,
\end{equation}
where the notation $({\rm l.h. E.})_n$ denotes 
additive terms that are linear and homogeneous 
in the Einstein tensor and its derivatives
of order $\leq n$. Form the tensor
\begin{equation}
\Omega_{ij}'=\Omega_{ij}-g_{ij}\Omega^k \mathstrut_k =
\nabla_0G_{ij}-2\nabla_{(i}G_{j)0}+g_{ij}\nabla^kG_{k0}+
({\rm l.h. E.})_0.
\end{equation}
{}From the twice-contracted Bianchi identities 
$\nabla^\mu G_{\mu \nu} \equiv 0$, we have
\begin{eqnarray}
\nabla^0G_{00}+\nabla^jG_{j0}\equiv 0, \\
\nabla^0G_{0i}+\nabla^jG_{ji}\equiv 0.
\end{eqnarray}
Compute $\nabla^j \Omega_{ij}'$, use the Ricci identity to
commute covariant derivatives, and use (53) to obtain
\begin{equation}
\Box G_{0i}+({\rm l.h. E.})_1 = 0.
\end{equation} 

But $G_{\alpha\beta}=0$ on $\Sigma_0$ by hypothesis and
$\dzeroh G_{\alpha \beta}=0$ (or $\nabla_0 G_{\alpha\beta}=0$) 
on $\Sigma_0$ from
(51), (52) and (53).  Hence $G_{0i}=0$ on $\Sigma_0\times{\rm I}$.
(The action of $\dzeroh$ on $G_{\alpha \beta}$ is defined because
in our basis, $G_{00}$ is a space scalar, $G_{0i}$ is a space
one-form, and $G_{ij}$ a space tensor.)

{}From $G_{0i}=0$ on $\Sigma_0\times{\rm I}$ and (51), it now
follows that $\dzeroh G_{ij}=0$; therefore, $G_{ij}=0$ on 
$\Sigma_0\times{\rm I}$ because it was zero on $\Sigma_0$.
Likewise, from (52), we infer that $G_{00}=0$ on $\Sigma_0\times{\rm I}$,
which completes the proof.

The passage to FOSH form of the Einstein equations proceeds similarly to that
for the Maxwell equations except for two points. One is technical: 
the non-trivial commutator (27) has to be used.  The other is conceptual: 
the necessity of using a generalized harmonic time-slicing condition. In
what follows, {\it we shall use simple harmonic slicing {\rm (44)} in the 
vacuum
case}; use of (45) only adds known terms and no new variables.
The key point about harmonic slicing $\dzero N+N^2 H=0$ is that when it
is combined with the Einstein equation $R_{00}=0$, i.e. 
\begin{equation}
\dzero H = - \bar\Delta N+K_{mk}K^{mk}N,
\end{equation}
we obtain a wave equation for $N$:
\begin{equation}
\Boxh N - (K_{ij}K^{ij}-H^2)N=0 .
\end{equation}

It is at this stage that the Hamiltonian constraint enters
explicitly into our system: see (36).
By applying $\bar \nabla_k$ to this equation we obtain a non-linear
wave equation for $a_k$. We may then reduce the
pair of wave equations involving $\Boxh K_{ij}$ and $\Boxh a_k$
each to flux-conservative FOSH form as we did previously for 
$\Box E_i$.
Furthermore, the new variables we must introduce to
carry this out are precisely the ones required to express the non-linear
right-hand-sides of these two wave equations in terms of the
defined variables, given the elimination of $\bar R_{ij}$ in these
equations using the standard Einstein equation of motion (33)
with $R_{ij}=0$.  ($\bar R_{ijkl}$ is to be regarded as eliminated from the
right-hand-side of the $\Boxh K_{ij}$ equation using (32): this key
step in obtaining FOSH form uses the fact of {\it three} space dimensions.
The $\Boxh K_{ij}$ equation itself has the same form in any Lorentzian 
(d+1) spacetime for the Einstein equations.)  These steps are executed
in \cite{CBY95,prl,CBY96,AY96}, 
but here we shall state the final results slightly more explicitly.

The new variables needed to reduce the $\Boxh K_{ij}$ and 
$\Boxh a_i$ wave equations are, respectively,
\begin{equation}
L_{0ij} \equiv N^{-1} \dzeroh K_{ij} ~~~~~~~~~~~~  
L_{kij}\equiv \bar\nabla_k K_{ij},
\end{equation}
\begin{equation}
a_{0i}\equiv N^{-1} \dzeroh a_i ~~~~~~~~~~~~~~~ a_{ji}\equiv\bar\nabla_j a_i.
\end{equation}
(In previous works we denoted $L_{0ij}$ by $L_{ij}$ and $L_{kij}$ by $M_{kij}$.)

Altogether, our unknowns up to this point in
the analysis are $N$, $g_{ij}$, $K_{ij}$,  $a_i$, $a_{0i}$, $a_{ji}$, 
$L_{0ij}$, and $L_{kij}$.
Note that the shift $\beta^i$ is not among the unknown variables.
Like $\phi(t,x^j)$ in the case of Maxwell's equations, $\beta^i(t,x^j)$ may
be regarded as {\it given} or {\it obtained} (as a function of $t$ and
$x^j$) in any manner that is mathematically compatible with the
Einstein equations in FOSH form.  
Obtaining $\beta^i(t,x^j)$ is needed in solving Einstein's equations, but
its ``manufacture'' is not a part of Einstein's equations
{\it per se}.

The reduction to first order form produces the following system
(in vacuum, with simple harmonic slicing)
\begin{equation}
\dzeroh N=-N^2H,
\end{equation} 
\begin{equation}
\dzeroh g_{ij}= -2N K_{ij},
\end{equation}
\begin{equation}
\dzeroh K_{ij}= NL_{0ij},
\end{equation}
\begin{equation}
\dzeroh L_{0ij}-N \bar\nabla^k L_{kij} = N {\cal J}_{ij} ,
\end{equation}
\begin{eqnarray}
\dzeroh L_{kij}-N\bar\nabla_k L_{0ij}=
N \left[ a_k L_{0ij} +2 L_{km(i} K_{j)}\mathstrut^m
+ 2 K_{m(i}L_{j)k}\mathstrut^m-2K_{m(i}L^m\mathstrut_{j)k} \right. \nonumber\\
\left. +2K_{m(i} \left(K_{j)}\mathstrut^m a_k +a_{j)}K_k\mathstrut^m
-a^mK_{j)k} \right) \right], 
\end{eqnarray}
\begin{equation}
\dzeroh a_i = N a_{0i} ,
\end{equation}
\begin{eqnarray}
\dzeroh a_{0i} - N \bar\nabla^k a_{ki}=
N \left[ H L_{ik}\mathstrut^k - 2L_{imk}K^{mk} +a^k \left(
a_{ki}-L_{0ik} + H K_{ik} - 2 K_{im}K^m\mathstrut_k \right) \right.\nonumber \\
\left.+a_i \left( H^2 +a^k a_k +2 a^k\mathstrut_k -2 K_{mk}K^{mk}\right)\right],
\end{eqnarray}
\begin{equation}
\dzeroh a_{ij}-N \bar\nabla_j a_{0i} = N \left[ a_k \left(
2L_{(ij)}\mathstrut^k - L^k\mathstrut_{ij}+2a_{(i}K_{j)}\mathstrut^k
-a^k K_{ij} \right) +a_j a_{0i} \right],
\end{equation}
where, in (62), $N{\cal J}_{ij}=J_{ij}$ with insertion of the
unknowns and elimination of spatial Riemann and Ricci tensors:
\begin{eqnarray}
{\cal J}_{ij} = 
K_{ij} \left( 2K_{mk}K^{mk}-3H^2  + a^k a_k \right) 
+ H \left(3 L_{0ij}  + 10 K_{ik} K^k\mathstrut_j 
+ 6 a_i a_j + 4 a_{ij}  \right) \nonumber \\
\hspace{-1.5cm}+ a_k \left( 3L^k\mathstrut_{ij}-4L_{(ij)}\mathstrut^k\right)
+4a_{(i}L_{j)k}\mathstrut^k 
 -\left(10 L_{0k(i}+ 16 K_{km} K^m\mathstrut_{(i}
+10 a_k a_{(i}+8 a_{k(i} \right) K_{j)}\mathstrut^k
 \\
+2g_{ij} K^{mk} \left(L_{0mk}- H K_{mk}+2K_{ml}K^l\mathstrut_k 
+a_ma_k+a_{mk} \right)
-g_{ij}H\left( K_{mk}K^{mk} - H^2 \right) . \nonumber
\end{eqnarray}
Note that the right-hand sides of (59)-(66) are polynomials in the unknowns
and in $g^{ij}$.  However, the reduction to first order form is not
quite complete because the spatial covariant derivatives on the
left-hand sides of (62), (63), (65), and (66) involve derivatives of
the unknown $g_{ij}$.  Two ways to complete the reduction are as follows.

One could introduce a known fiducial torsion-free connection $\bar F^i_{jk}
(t,x^l)$ on $\Sigma_0\times {\rm I}$ whose associated covariant derivative
operator is (say) $\bar D_i$.  Then define the tensor
\begin{equation}
S^i_{jk}=\bar \Gamma^i_{jk}-\bar F^i_{jk}(t,x^l)
\end{equation}
as a new unknown whose equation of motion is
\begin{equation}
\dzeroh S^i_{jk}=\dzeroh \bar \Gamma^i_{jk}-\dzeroh \bar F^i_{jk}(t,x^l),
\end{equation}
where
\begin{equation}
\dzeroh \bar \Gamma^i_{jk}=-N[2L_{(jk)}\mathstrut^i- L^i\mathstrut_{jk}
+2 a_{(j}K_{k)}\mathstrut^i-a^i K_{jk}].
\end{equation}
The equation obtained from combining (69) and (70) can be appended to
(59)-(66) to form the complete first order system.  The operators
$\bar\nabla_i$ are replaced by relations of the form
\begin{equation}
\bar\nabla_k T_{ij}=\bar D_k T_{ij} -T_{lj} S^l_{ik} -T_{il}S^l_{jk}.
\end{equation}

Another method that can be used to complete the reduction is simply
to append the relation (70) and to rewrite the $\bar \nabla_i$'s
in terms of ordinary derivatives: {\it all} the equations of the
system are still spatially covariant though some will have, of course,
individual terms which are not tensors.  This is true, perhaps not
obviously, of (70) also.  One has to recall that
\begin{eqnarray}
{\cal L}_\beta \bar \Gamma^i_{jk} &=&(\beta^l \partial_l \bar \Gamma^i_{jk}
-\bar \Gamma^l_{jk} \partial_l \beta^i 
+\bar \Gamma^i_{lk} \partial_j \beta^l
+\bar \Gamma^i_{jl} \partial_k \beta^l)  
+\partial_j \partial_k \beta^i \\
&\equiv& \bar\nabla_j \bar\nabla_k \beta^i - \bar R^i\mathstrut_{kjl}\beta^l
\nonumber 
\end{eqnarray}
is a spatial tensor.

In our system, (62) and (65) are wave equations and (63) and (66) are,
respectively, their ``adjoint'' equations.  All the other equations
are definitions except for the slicing equation (59) and the equation
for the evolution of the connection.  This system is {\it symmetrizable}
upon multiplication by a symmetric positive definite matrix consisting of 1's
down the main diagonal and two sub-matrices of the form (17) centered
on the main diagonal to handle the two wave equations and their adjoints.
The unknowns can be expressed by
\begin{equation}
{\bf u}^T=({\bf\bar g}, {\bf K}, {\bf L_0}, {\bf L_1}, {\bf L_2},
{\bf L_3}, N, {\bf a}, {\bf a_0}, {\bf a_1}, {\bf a_2}, {\bf a_3},
{\bf \bar \Gamma}),
\end{equation}
where a trailing pair of indices is suppressed from
${\bf\bar g},\ {\bf K},$ and ${\bf L}_{\mu}$, a trailing vector index
from ${\bf a}$ and ${\bf a}_\mu$ and all of the indices from ${\bf
\bar\Gamma}$.  Therefore, we have constructed a 
complete FOSH system.  The structure of the left-hand sides has the 
``flux-conservative'' form possible for curved space and curvilinear
coordinates.

For completeness, we state explicitly the initial data for
the FOSH system (59)-(66) and (70), given $\beta^i(t,x^j)$.  
These data are consistent with those of (47), (48)
and (49) with $f=0$.
The lapse function $N>0$ is chosen freely, which implies the
initial acceleration $a_i = \partial_i \log N$.  The metric
$g_{ij}$ and extrinsic curvature $K_{ij}$ satisfy the usual Gauss-Codazzi
constraints $C=0$ and $C_i=0$ given by (37C). From $g_{ij}$ and
$K_{ij}$ we obtain the initial values of $\bar \Gamma^i_{jk}$
and of $L_{ijk}=\bar\nabla_i K_{jk}$, as well as those of 
$a_{ji}= \bar\nabla_j \partial_i \log N$.  The initial values of
$L_{0ij}=N^{-1} \dzeroh K_{ij}$ are given by $R_{ij}=0$
in (33) and those of $a_{0i}=N^{-1} \dzeroh a_i$ by
$-(\partial_i H + H \partial_i \log N)$, where the
simple harmonic slicing condition $\partial_0 N +N^2 H=0$
was used in the latter. 

Upon inspection, one sees that all fields evolve along
the $\partial_0$ axis except those obtained from the wave equations
and their adjoints.
Dividing these equations through by $N$ (to obtain proper time),
just as in the case of Maxwell's theory we obtain characteristic
fields of the form, for the choice $x^j=x^1$,
\begin{eqnarray}
&&{1\over \sqrt{2}} \alpha [L_{0ij} \pm \alpha^{-1} (L_{1ij} -v^a
L_{aij})], \\
&&{1\over \sqrt{2}} \alpha [a_{0i} \pm \alpha^{-1} (a_{1i} - v^a
a_{ai})].
\end{eqnarray}
These fields propagate at light speed along the cone, and  they have
the dimensions of curvature, as evident from (29) and (30).  At a given 
event, the ``electric''   or ``tidal'' parts of the Riemann
tensor ($N^{-2} R_{0i0j}$) are constructed
from contributions from these fields, which arrive
from along the past cone, and from additional contributions which
arrive from along the past timelike 
curve orthogonal to the slice $\Sigma_t$ of the
given foliation in which the event lies.

By writing (74) and (75) in the equivalent forms
\begin{eqnarray}
&&{1\over \sqrt{2}} \alpha [N^{-1}\dzeroh \pm \alpha^{-1} (\bar\nabla_1 -v^a
\bar\nabla_a)]K_{ij}, \\
&&{1\over \sqrt{2}} \alpha [N^{-1}\dzeroh \pm \alpha^{-1} (\bar\nabla_1 -v^a
\bar\nabla_a)]a_i,
\end{eqnarray}
we see that the ``characteristic fields''
 are differential extrinsic curvature and 
differential acceleration, with differences taken along the past
cone of an event using the many pairs of null generators corresponding to
advanced and retarded times for all spatial directions
in $\Sigma_t$ with origin at the event.

Finally, we remark that net matter contribution to the wave equations
(62) and (65) are included by adding to the right-hand sides the
quantities
\begin{eqnarray}
\Theta_{ij}&\equiv& N \tau_{ij}= N\kappa [-N^{-1}\dzeroh \rho_{ij} +
2N^{-1}\bar\nabla_{(i} \rho_{j)0} + 
K_{ij}(2N^{-2}\rho_{00}+  \rho^k\mathstrut_k)  \\
&&\hspace{1cm} 
+2  H \rho_{ij} - 6  \rho_{k(i} K_{j)}\mathstrut^k 
+g_{ij}(2 K^{mk}\rho_{mk}  -H \rho^k\mathstrut_k
-H N^{-2}\rho_{00})], \nonumber \\
\Phi_i&\equiv& N\mu_i = -N\kappa  \left[ \rho_{ik} a^k
+N^{-2} \bar \nabla_i \rho_{00} \right] . 
\end{eqnarray}
Note that the detailed forms of (78) and (79) depend
on which Einstein equations we have used to display
the right hand sides of (62) and (65).

\section{Higher Order Formulation}

We may ask whether a wave equation form of the Einstein evolution equations 
can be found in which both the shift {\it and the lapse}
can be freely specified.  The answer is yes \cite{AACBYprep}, but
we will not give a full description here.  

Form the expression
\begin{equation}
\Lambda_{ij} \equiv \dzeroh \Omega_{ij} +\bar\nabla_i \bar\nabla_j R_{00}
\equiv \dzeroh \dzeroh R_{ij} - 2 \dzeroh \bar\nabla_{(i} R_{j)0}
+ \bar\nabla_i \bar\nabla_j R_{00}.
\end{equation}
This gives a system of the form
\begin{equation}
\Boxh\dzeroh K_{ij}= [ {\rm terms~with~derivatives~of~}
{\bf K} {\rm~of~order~} \leq 2 {\rm~and~of~} {\bar {\bf g}} \rm{~of~
order~} \leq 3 ],
\end{equation}
\begin{eqnarray}
\dzeroh g_{ij} &=& -2N K_{ij}, 
\end{eqnarray}
that has no slicing term analogous to $S_{ij}$.  It
has characteristics along the light cone and $\partial_0$.
It is not
quasi-diagonal because of 
third order derivatives of $g_{ij}$ 
arising in (81) from $\bar\nabla^2 \bar \Delta N$.
This system is hyperbolic non-strict \cite{LeOh}.
On the other hand, the
fourth order equation for $g_{ij}$ deduced from (81) and
(82) is also not strictly hyperbolic
because of the appearance of $\dzero^2$ in its
principal operator ($\dzero^2 \Box$).  It seems that the system, 
in its present form, cannot be written as a first order
symmetrizable system;
but it is potentially
very useful, especially in perturbation theory, because the lapse
and the shift need not be perturbed or satisfy any particular
gauge conditions.  The system (81) and (82) can be proven to be
equivalent to the full Einstein equations using the twice-contracted
Bianchi identities and the appropriate Cauchy data: $G_{\alpha \beta}=0$
and $\Omega_{ij}=0$ on $\Sigma_0$.  This system can be 
directly adapted for the 
study of the non-linear perturbative
regime of high-frequency wave propagation \cite{YCB1,YCB2}. 

\section{Concluding Remarks}

One of the numerous attractive reasons for writing general relativity
in flux-conservative FOSH form, as emphasized recently in 
\cite{BMSS}, is to facilitate emulating the successful strategy
applied in computational fluid dynamics, rather
than tailoring the form of Einstein's equations to suit each
particular application.  However, in fluid dynamics in 
flux-conservative FOSH form, the only characteristics that
can appear are the physical ones associated with the ``sound
cone'' or ``flow lines,'' because there are no analogs of the spatial coordinate
and slicing ``gauges'' of general relativity that give
rise to constraints and can lead to unphysical characteristics. 
In this paper, by modifying
the equations of motion of Einstein's theory in a specific
way ($R_{ij}\rightarrow \Omega_{ij}$), and by choosing
time slicings in which the lapse function obeys a causal wave
equation, we have incorporated full spatial coordinate freedom
while assuring that only physical characteristics are present. 
Among other things, this means that causal boundary conditions can
be imposed naturally on bounded regions of spacetime. 
Gravitational physics and its mathematical
expression have been smoothly melded.

Acknowledgements.
A.A., A.A., and J.W.Y. were supported by National Science Foundation
grants PHY-9413207 and PHY 93-18152/ASC 93-18152 (ARPA supplemented).
They thank Gregory Cook and Mark Scheel for helpful correspondence.

\end{document}